\documentclass[
reprint,
superscriptaddress,
amsmath,amssymb,
aps, prb,
aip,rsi,
floatfix
]{revtex4-2}

\usepackage{graphicx}
\usepackage{dcolumn}
\usepackage{bm}

\usepackage{filecontents}

\begin{document}

\preprint{APS/123-QED}

\author{S. Dietrich}%
\email{scott.dietrich@villanova.edu}
\affiliation{Physics Department, Villanova University, Villanova, PA 19089, USA}

\title{Hall signal-dominated microwave transmission through graphene-loaded waveguides}

\author{A. Akbari-Sharbaf} 
\affiliation{Physics Department, Villanova University, Villanova, PA 19089, USA}

\author{J.H. Edgar} 
\affiliation{Department of Chemical Engineering, Kansas State University, Manhattan, KS 66506, USA}

\author{A. Roubos} 
\author{M.L. Freeman}
\affiliation{Physics Department, Florida State University, Tallahassee, Florida 32306, USA}

\author{L.W. Engel}
\affiliation{Physics Department, Florida State University, Tallahassee, Florida 32306, USA}
\affiliation{National High Magnetic Field Laboratory, Tallahassee, Florida 32310, USA}

\date{\today}

\begin{abstract}
Microwave transmission line spectroscopy is used to observe the integer quantum Hall effect in two samples of monolayer graphene with different geometries that are resistively-coupled to a coplanar waveguide. We find plateaus in transmitted power that do not vary significantly with microwave frequency but are significantly different for two samples due to their shape. With each drop in transmitted power corresponding to an additional quantum Hall edge mode that short the transmission line to ground, these well-known quanta of conductance allow us to calibrate the sensitivity of the devices. One sample with short contact regions matched the sensitivity expected when considering only the quantum Hall conductance of $\nu e^2/h$; another sample with long contact regions demonstrated a nearly three-fold enhancement in sensitivity. We model this result with a purely resistive circuit that introduces an additional resistance to explain the increased sensitivity.
\end{abstract}

\maketitle

\section{Introduction}

The phase diagrams of 2D electron systems (2DES) in high magnetic field are a complicated interweaving of vastly different phases in the quantum Hall (QH) regime: fractional Quantum Hall effect liquids, Wigner crystals, bubbles, stripes, and more \cite{Wang2008,Zhang2007,Knoester2016,DaSilva2016}.  There are many ways to study the quantum Hall effects at high magnetic field and low temperatures. Traditional Hall bar or Corbino ring architectures yield important information, including the size of energy gaps and signs of electron crystallization. DC transport using Hall bar and Corbino ring geometries are two ways to simplify the measurements of voltages along a sample, taking advantage of Cartesian and plane polar symmetries respectively. Owing to their simplicity, these geometries remain important for investigating the QH effect and other novel effects in 2DES \cite{Dean2013, Zeng2019,Li2021,Chen2021}. Variations from these highly-symmetric cases need more advanced methods of analysis. The van der Pauw geometry, typically four contacts placed along the circumference of a circular disc or at the corners of a square, has been a convenient way to measure anisotropic effects \cite{Lilly1999} as well as Hall and longitudinal coefficients of isotropic samples in magnetic field. However, there has been a significant amount of work since then to extend the theory and improve calculation efficiency in practical devices \cite{deMey1983}. In fact, a descendant of van der Pauw geometry known as a sunflower is a convenient way to enable angle-resolved transport \cite{Zhang2024} for materials with complicated anisotropy.

Other methods are often utilized to gain additional insight into the nature of the many competing electronic phases of 2DES. Some of these methods are motivated by minimizing the need for contacts in materials where Ohmic contacts are difficult to achieve \cite{Cui2017,Tomarken2019, Gustafsson2018, Jin2021} while others lend additional information that is not easily accessible by transport measurements. Microwave transmission line spectroscopy (MWTS) has been a particularly important tool for understanding the competition of electron solid phases in 2DES hosted in GaAs/AlGaAs quantum wells \cite{Zhu2010, Hatke2014, Chen2006}. All of these examples use coplanar waveguides (CPWs) which are capacitively coupled to the 2DES. This architecture leads to a measurement of the longitudinal conductivity and is of value for studying the pinning modes of electron solids \cite{Andrei1988}. Understanding such phases are a large motivation for utilizing alternatives to DC transport, where some of these phases show up as the re-entrant integer QH effect \cite{Liu2012, Chen2019}. While MWTS spectroscopy has made a significant impact on the study of GaAs/AlGaAs quantum wells, it has not yet had similar success in the area of van der Waals (vdW) materials. This is primarily due to the low sensitivity in capacitively-coupled CPWs that come from the small areas of high-quality, single crystal exfoliated samples in standard 50 $\Omega$ measuring systems and from the larger disorder of large-area CVD-grown samples which could be patterned with long narrow CPWs.

In this work, we study monolayer graphene that is directly contacted to a CPW and observe that the transmitted microwave power is mainly determined by the Hall rather than longitudinal conductivity. For one sample, we also observe almost a tripling of the expected sensitivity, which we model with an additional resistance along the center line of the CPW.

\section{Experimental Setup}

\begin{figure*}
\begin{center}
    \includegraphics[width = \linewidth]{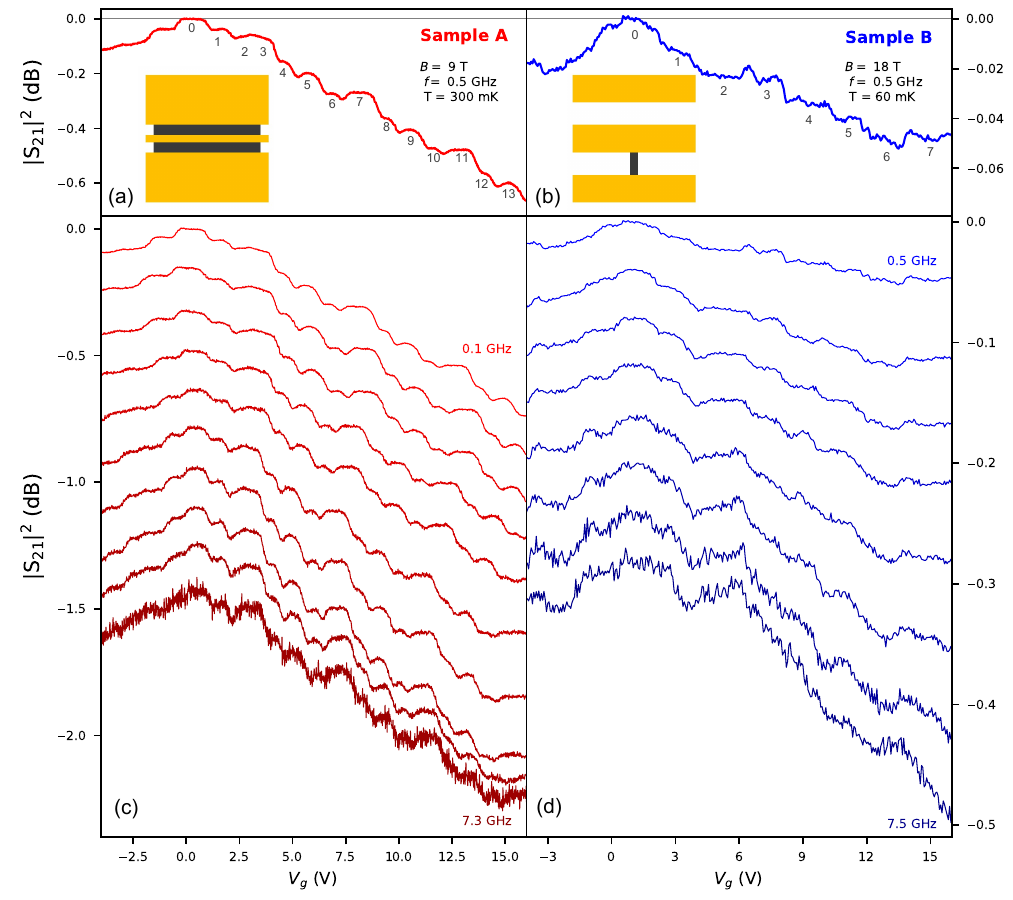}
\end{center}
\caption{\label{main} (a,b) Low frequency transmitted power as a function of gate voltage $V_g$ for Samples A (a) and B (b). $\nu$ values are listed below each plateau. Sample A shows steps between QH plateaus that are over twice as large as those in Sample B. The insets show the geometries of the graphene layers. (c,d) Lower panels demonstrate how the plateau structure lacks any significant frequency-dependence for frequencies up to a few GHz. Curves in all panels have been normalized to the value of the $\nu$=0 plateau. Curves for Sample A (c) have frequency steps of 0.8 GHz and are offset by 160 mdB. Curves for Sample B (d) have frequency steps of 1.0 GHz and are offset by 40 mdB. }
\end{figure*}

In MWTS, incoming MW radiation travels down 50 $\Omega$ impedance lines to on-chip CPWs that reduce down to appropriate dimensions for the graphene heterostructure while retaining impedance matching. The CPW architecture allows for broadband measurements of the absorption and phase-shift from a graphene sample using vector a network analyzer (VNA). In this work, we focus on the transmission coefficient of power, $|S_{21}|^2$. 

Our resistively-coupled CPWs in graphene have ohmic contact to the graphene layer at both the center line and grounding planes. The additional contact between the center line and the graphene is the main difference from capactively-coupled structures. Heterostructures are formed by dry transfer assembly on intrinsic Si wafers with a 285 nm oxide layer. Cr/Au (3/80 nm) CPWs are deposited over the heterostructures. The grounding planes and center conductor make direct contact to the graphene layer via trenches made in the top hBN layer. It is designed such that the center-line inductance per unit length $L$ and the center-line-to-ground-plane capacitance per unit length $C$ provide $Z=\sqrt{L/C}=$ 50 $\Omega$.  The graphene adds a conductivity $G_s$ that acts as a direct shunt to ground. The density of charge carriers in the graphene layer are controlled by a voltage $V_g$ applied to a thin graphite back gate relative to the two grounding planes on either side of the center line.

Data in this study comes from two geometrically different samples. They are characterized by the dimension of the graphene that has a width, $s$, and length along the CPW, $\ell$. The first, referred to as the Sample A, has $s\times\ell=$ 3 $\times$ 33 $\mu$m, while the second, Sample B, measures 20 $\times$ 3 $\mu$m. The two geometries are illustrated in the inset of Figure \ref{main} (a) and (b). Both samples have similar hBN layers separating the graphene layer from the gate, 78 nm for Sample A and  84 nm for Sample B. The CPWs of the devices have center-line widths of 1.0 and 33.3 $\mu$m and room temperature resistances of 110 and 90 $\Omega$ for Samples A and B respectively. Measurements of Sample A occurred at a magnetic field of 9 T and at a temperature of 300 mK. Measurements of Sample B occurred at a magnetic field of 18 T and at a bath temperature of 60 mK.

\section{Experimental Results}

\begin{figure}
        \includegraphics[width=\linewidth]{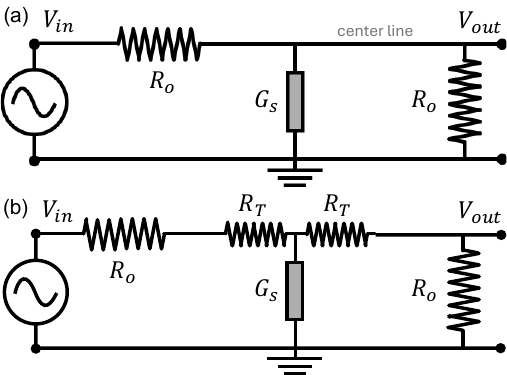}
        \caption{(a) Simple model of a two-point quantum Hall measurement where the top horizontal wire represents the center line of the CPW and the bottom horizontal wire is the two grounding planes. $G_s$ is the conductance of the graphene. (b) T-network model that accounts for the additional center-line resistance $R_T$ on either side of the sample.}
    \label{simple}
\end{figure}

Rather than the longitudinal conductivity dominating variations in transmitted power as seen in capacitivitely-coupled samples, our resistively-coupled graphene-loaded CPWs show major contributions from the Hall conductivity in $|S_{21}|^2$, resembling two point DC conductance of a QH sample. This can be clearly seen in Figure \ref{main}a and \ref{main}b where the plateaus are as expected for the integer QH effects with the filling factors $\nu$ indicated. Only the integer QH is observed, likely due to reduced sample quality from close proximity of contact and the single-graphite gating structure that leaves the graphene exposed to surface contaminants. However, there are two notable features in the sequence of plateaus between the two samples. Firstly, jumps between $\nu=$ 2 $\rightarrow$ 3, 6 $\rightarrow$ 7, and 10 $\rightarrow$ 11 are absent in both samples. Gate geometry calculations confirm the location of these plateaus. We note that these occur distinctly at $\nu=4n+2$ associated with the four-fold degeneracy of monolayer graphene, but a detailed explanation of this observation is beyond the scope of this manuscript. 

The other feature is that the size of the jumps vary greatly depending on the geometry of the sample. These plateaus should allow us to calibrate the sensitivity of the device since they each correspond to an additional edge channel conductance, which is $G_o=e^2/h$ = 38.7 $\mu$S when considering full degeneracy breaking. Figure \ref{main}a and \ref{main}b demonstrate how the drops in transmitted power are much stronger for Sample A than Sample B. On average, Sample A shows drops of 36 mdB per Landau level while Sample B shows drops of 14 mdB. These correspond to sensitivities of -1244 dB/S and -179 dB/S respectively. For both samples, the plateau structure is independent of microwave frequency, as shown in Figure \ref{main}c and \ref{main}d. This motivates using a simple resistive model of a two-point measurement of the Hall conductance. This investigation will focus on the frequency-independent variations in the sensitivity differences between the two samples. Figure \ref{schematics}a shows how the center conductors are shunted to the ground by the graphene. The two $R_o$ = 50 $\Omega$ resistors are equivalent to the MW source side (left) and the detector side (right) of the VNA. The conductivity of the graphene is considered to be $G_s=N\nu G_o$ when there are either $N =$ 1 or 2 loaded slots in the CPW and since $\sigma_{xx}=0$ in the QH plateaus.

The sensitivity, $\eta$, can be written as
\begin{align}
    \label{sensitivity}
    \displaystyle\eta=\frac{\Delta P\text{(dB)}}{\Delta G}=\frac{10\log_{10}{\left(P_1/P_0\right)}}{G_o}
\end{align}
where $P_\nu=V_{out}^2/R_o$ indicates the power transmitted through the CPW when the sample is in the $\nu$ QHE state and we assume a $\Delta G=G_o$ for each $\Delta\nu=1$ transition between Landau levels. To determine how these powers depend on the circuit elements, we use Kirchhoff's junction and loop rules to express the the voltage at the analyzer $V_{out}$ in terms of the incident voltage $V_{in}$. For the simple model of Figure 2a,
\begin{align}
    V_{out}&=\frac{1}{2+N\nu G_oR_o}\cdot V_{in}
    \label{schematics}
\end{align}
which can be used with Equation \ref{sensitivity} to calculate the sensitivity for the simple model. It predicts sensitivities of -434 dB/S for Sample A and -217 dB/S for Sample B.  These sensitivities correspond to 16.8 and 8.4 mdB drops in transmitted power per Landau level. This model closely predicts the sensitivity of Sample B. In fact, Sample B has a sensitivity 38 dB/S lower, an effect likely due to a small amount of contact resistance to the graphene. Conversely, this simple model greatly underestimates the sensitivity of Sample A. This significant, nearly three-fold, deviation from the observed sensitivity is a shortcoming of this simple model. Any additional resistance of the sample (i.e. contact resistance) would only \textit{decrease} the sensitivity as seen in Sample B. Thus, we model the increased sensitivity as due to an additional resistance along the center line of the CPW. The model is shown in Figure \ref{schematics}b, where the graphene forms a T-network with two resistors $R_T$. These resistances provide the increased sensitivity of the CPW.  A general expression for the transmitted power when the graphene is in the QH effect state at filling $\nu$ is given by
\begin{align}
    \label{Pnu}
    P_\nu&=(\gamma_\nu V_{in})^2/R_o\\
    \gamma_\nu &= \frac{R_o}{2(R_o+R_T)+N\nu G_o(R_o+R_T)^2}\label{modelEQN}
\end{align}

\begin{figure}
        \includegraphics[width=\linewidth]{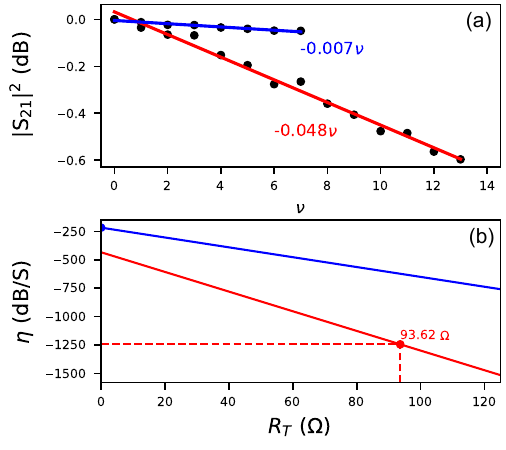}
        \caption{(a) Transmitted power at each Landau level with filling factor $\nu$ for Sample A (red) and Sample B (blue) at $f=$ 0.5 GHz. (b) Dependence of the sensitivity, $\eta$, on $R_T$ based on the model of Equations 3 and 4. Colors match the plot above for the two samples. Dots with dashed lines mark the $\eta$ and $R_T$ of each sample.}
    \label{models}
\end{figure}

Figure \ref{models}a shows the transmitted power for each Landau level. The slopes of the linear regressions give the $\Delta P$ due to changes in sample conductance of $G_o$, allowing us to extract the sensitivity according to Equation \ref{sensitivity}. Figure \ref{models}b demonstrates how the sensitivity varies with the additional center-line resistance according to Equation \ref{modelEQN}. For the relevant values of $R_T$ the sensitivity increases nearly linearly and the sensitivity roughly doubles with an additional 50 $\Omega$ resistance on either side of the sample. We also recover the simpler model's sensitivities of -217 dB/S and -434 dB/S for the singly- and doubly-loaded CPW as $R_T$ goes to zero as expected. After extracting the sensitivity of -1244 dB/S from the slope of $|S_{21}|^2$ in Figure \ref{models}a, this model predicts an $R_T$ value of 93.6 $\Omega$ for Sample A.


\section{Concluding Remarks}

The T-network model of Figure \ref{models}b reproduces the observed sensitivity for Sample A as calculated for the known Hall resistance of the QH plateaus, with $R_T$ of 93.6 $\Omega$.  While this simple model reproduces the observed sensitivity in Sample A, by presenting a higher impedance than 50 $\Omega$ to the samples, the real situation in the sample is likely different. The Au/Cr center lines of the CPWs should not have sufficient resistance at low temperature to explain the above $R_T$ values from the model. The origin of the sensitivity enhancement relative to the model of Figure \ref{models}a  with  $G_s=N\nu G_0$ on QH plateaus is more likely an effect of the finite size of the contacts. Calculations of such effects can require considerable complexity \cite{deMey1983}.

In sum we have measured two graphene samples with large disorder by modern standards \cite{Zeng2019} at microwave frequency and found almost no frequency dependence. The microwave sensitivity to the two-point conductance, $G_s$, of the graphene calculated from the slope of $|S_{21}|^2$ is compared to what is calculated taking $G_s=\nu e^2/h$ on QH plateaus in a 50 $\Omega$ system. Sample B, which had short contact regions, matched these expectations. Sample A, which had long contact regions, showed a nearly three-fold enhancement and can be modeled with a simple T-network circuit.


\begin{acknowledgments}
This material is based upon work supported by the National Science Foundation under Grant No. DMR-1943389. Part of the work was carried out in part at the Singh Center for Nanotechnology, which is supported by the NSF National Nanotechnology Coordinated Infrastructure Program under grant NNCI-2025608. A portion of this work was performed at the National High Magnetic Field Laboratory, which is supported by National Science Foundation Cooperative Agreement No. DMR-2128556 and the State of Florida.
\end{acknowledgments}


\begin{thebibliography}{21}%
\makeatletter
\providecommand \@ifxundefined [1]{%
 \@ifx{#1\undefined}
}%
\providecommand \@ifnum [1]{%
 \ifnum #1\expandafter \@firstoftwo
 \else \expandafter \@secondoftwo
 \fi
}%
\providecommand \@ifx [1]{%
 \ifx #1\expandafter \@firstoftwo
 \else \expandafter \@secondoftwo
 \fi
}%
\providecommand \natexlab [1]{#1}%
\providecommand \enquote  [1]{``#1''}%
\providecommand \bibnamefont  [1]{#1}%
\providecommand \bibfnamefont [1]{#1}%
\providecommand \citenamefont [1]{#1}%
\providecommand \href@noop [0]{\@secondoftwo}%
\providecommand \href [0]{\begingroup \@sanitize@url \@href}%
\providecommand \@href[1]{\@@startlink{#1}\@@href}%
\providecommand \@@href[1]{\endgroup#1\@@endlink}%
\providecommand \@sanitize@url [0]{\catcode `\\12\catcode `\$12\catcode `\&12\catcode `\#12\catcode `\^12\catcode `\_12\catcode `\%12\relax}%
\providecommand \@@startlink[1]{}%
\providecommand \@@endlink[0]{}%
\providecommand \url  [0]{\begingroup\@sanitize@url \@url }%
\providecommand \@url [1]{\endgroup\@href {#1}{\urlprefix }}%
\providecommand \urlprefix  [0]{URL }%
\providecommand \Eprint [0]{\href }%
\providecommand \doibase [0]{https://doi.org/}%
\providecommand \selectlanguage [0]{\@gobble}%
\providecommand \bibinfo  [0]{\@secondoftwo}%
\providecommand \bibfield  [0]{\@secondoftwo}%
\providecommand \translation [1]{[#1]}%
\providecommand \BibitemOpen [0]{}%
\providecommand \bibitemStop [0]{}%
\providecommand \bibitemNoStop [0]{.\EOS\space}%
\providecommand \EOS [0]{\spacefactor3000\relax}%
\providecommand \BibitemShut  [1]{\csname bibitem#1\endcsname}%
\let\auto@bib@innerbib\@empty
\bibitem [{\citenamefont {Wang}\ \emph {et~al.}(2008)\citenamefont {Wang}, \citenamefont {Iyengar}, \citenamefont {Fertig},\ and\ \citenamefont {Brey}}]{Wang2008}%
  \BibitemOpen
  \bibfield  {author} {\bibinfo {author} {\bibfnamefont {J.}~\bibnamefont {Wang}}, \bibinfo {author} {\bibfnamefont {A.}~\bibnamefont {Iyengar}}, \bibinfo {author} {\bibfnamefont {H.~A.}\ \bibnamefont {Fertig}},\ and\ \bibinfo {author} {\bibfnamefont {L.}~\bibnamefont {Brey}},\ }\bibfield  {title} {\bibinfo {title} {Phase diagram for quantum {Hall} states in graphene},\ }\href@noop {} {\bibfield  {journal} {\bibinfo  {journal} {Phys. Rev. B}\ }\textbf {\bibinfo {volume} {78}},\ \bibinfo {pages} {165416} (\bibinfo {year} {2008})}\BibitemShut {NoStop}%
\bibitem [{\citenamefont {Zhang}\ and\ \citenamefont {Joglekar}(2007)}]{Zhang2007}%
  \BibitemOpen
  \bibfield  {author} {\bibinfo {author} {\bibfnamefont {C.-H.}\ \bibnamefont {Zhang}}\ and\ \bibinfo {author} {\bibfnamefont {Y.~N.}\ \bibnamefont {Joglekar}},\ }\bibfield  {title} {\bibinfo {title} {{Wigner} crystal and bubble phases in graphene in the quantum {{Hall}} regime},\ }\href@noop {} {\bibfield  {journal} {\bibinfo  {journal} {Phys. Rev. B}\ }\textbf {\bibinfo {volume} {75}},\ \bibinfo {pages} {245414} (\bibinfo {year} {2007})}\BibitemShut {NoStop}%
\bibitem [{\citenamefont {Knoester}\ \emph {et~al.}(2016)\citenamefont {Knoester}, \citenamefont {Papic},\ and\ \citenamefont {Morais~Smith}}]{Knoester2016}%
  \BibitemOpen
  \bibfield  {author} {\bibinfo {author} {\bibfnamefont {M.~E.}\ \bibnamefont {Knoester}}, \bibinfo {author} {\bibfnamefont {Z.}~\bibnamefont {Papic}},\ and\ \bibinfo {author} {\bibfnamefont {C.}~\bibnamefont {Morais~Smith}},\ }\bibfield  {title} {\bibinfo {title} {Electron-solid and electron-liquid phases in graphene},\ }\href@noop {} {\bibfield  {journal} {\bibinfo  {journal} {Phys. Rev. B}\ }\textbf {\bibinfo {volume} {93}},\ \bibinfo {pages} {155141} (\bibinfo {year} {2016})}\BibitemShut {NoStop}%
\bibitem [{\citenamefont {DaSilva}\ \emph {et~al.}(2016)\citenamefont {DaSilva}, \citenamefont {Jung},\ and\ \citenamefont {MacDonald}}]{DaSilva2016}%
  \BibitemOpen
  \bibfield  {author} {\bibinfo {author} {\bibfnamefont {A.~M.}\ \bibnamefont {DaSilva}}, \bibinfo {author} {\bibfnamefont {J.}~\bibnamefont {Jung}},\ and\ \bibinfo {author} {\bibfnamefont {A.~H.}\ \bibnamefont {MacDonald}},\ }\bibfield  {title} {\bibinfo {title} {Fractional {Hofstadter} states in graphene on hexagonal boron nitride},\ }\href@noop {} {\bibfield  {journal} {\bibinfo  {journal} {Phys. Rev. Lett.}\ }\textbf {\bibinfo {volume} {117}},\ \bibinfo {pages} {036802} (\bibinfo {year} {2016})}\BibitemShut {NoStop}%
\bibitem [{\citenamefont {Dean}\ \emph {et~al.}(2013)\citenamefont {Dean}, \citenamefont {L.~Wang}, \citenamefont {Forsythe}, \citenamefont {Ghahari}, \citenamefont {Gao}, \citenamefont {Katoch}, \citenamefont {Ishigami}, \citenamefont {Moon}, \citenamefont {Koshino}, \citenamefont {Taniguchi}, \citenamefont {Watanabe}, \citenamefont {Shepard},\ and\ \citenamefont {Kim}}]{Dean2013}%
  \BibitemOpen
  \bibfield  {author} {\bibinfo {author} {\bibfnamefont {C.~R.}\ \bibnamefont {Dean}}, \bibinfo {author} {\bibfnamefont {P.~M.}\ \bibnamefont {L.~Wang}}, \bibinfo {author} {\bibfnamefont {C.}~\bibnamefont {Forsythe}}, \bibinfo {author} {\bibfnamefont {F.}~\bibnamefont {Ghahari}}, \bibinfo {author} {\bibfnamefont {Y.}~\bibnamefont {Gao}}, \bibinfo {author} {\bibfnamefont {J.}~\bibnamefont {Katoch}}, \bibinfo {author} {\bibfnamefont {M.}~\bibnamefont {Ishigami}}, \bibinfo {author} {\bibfnamefont {P.}~\bibnamefont {Moon}}, \bibinfo {author} {\bibfnamefont {M.}~\bibnamefont {Koshino}}, \bibinfo {author} {\bibfnamefont {T.}~\bibnamefont {Taniguchi}}, \bibinfo {author} {\bibfnamefont {K.}~\bibnamefont {Watanabe}}, \bibinfo {author} {\bibfnamefont {K.~L.}\ \bibnamefont {Shepard}},\ and\ \bibinfo {author} {\bibfnamefont {J.~H. .~P.}\ \bibnamefont {Kim}},\ }\bibfield  {title} {\bibinfo {title} {{Hofstadter}’s butterfly and the fractal quantum {Hall} effect in moiré superlattices},\ }\href@noop {} {\bibfield
  {journal} {\bibinfo  {journal} {Nature}\ }\textbf {\bibinfo {volume} {497}},\ \bibinfo {pages} {598–602} (\bibinfo {year} {2013})}\BibitemShut {NoStop}%
\bibitem [{\citenamefont {Zeng}\ \emph {et~al.}(2019)\citenamefont {Zeng}, \citenamefont {Li}, \citenamefont {Dietrich}, \citenamefont {Ghosh}, \citenamefont {Watanabe}, \citenamefont {Taniguchi}, \citenamefont {Hone},\ and\ \citenamefont {Dean}}]{Zeng2019}%
  \BibitemOpen
  \bibfield  {author} {\bibinfo {author} {\bibfnamefont {Y.}~\bibnamefont {Zeng}}, \bibinfo {author} {\bibfnamefont {J.}~\bibnamefont {Li}}, \bibinfo {author} {\bibfnamefont {S.}~\bibnamefont {Dietrich}}, \bibinfo {author} {\bibfnamefont {O.}~\bibnamefont {Ghosh}}, \bibinfo {author} {\bibfnamefont {K.}~\bibnamefont {Watanabe}}, \bibinfo {author} {\bibfnamefont {T.}~\bibnamefont {Taniguchi}}, \bibinfo {author} {\bibfnamefont {J.}~\bibnamefont {Hone}},\ and\ \bibinfo {author} {\bibfnamefont {C.}~\bibnamefont {Dean}},\ }\bibfield  {title} {\bibinfo {title} {High-quality magnetotransport in graphene using the edge-free corbino geometry},\ }\href@noop {} {\bibfield  {journal} {\bibinfo  {journal} {Physical Review Letters}\ }\textbf {\bibinfo {volume} {122}} (\bibinfo {year} {2019})}\BibitemShut {NoStop}%
\bibitem [{\citenamefont {Li}\ \emph {et~al.}(2021)\citenamefont {Li}, \citenamefont {Dietrich}, \citenamefont {Forsythe}, \citenamefont {Taniguchi}, \citenamefont {Watanabe}, \citenamefont {Moon},\ and\ \citenamefont {Dean}}]{Li2021}%
  \BibitemOpen
  \bibfield  {author} {\bibinfo {author} {\bibfnamefont {Y.}~\bibnamefont {Li}}, \bibinfo {author} {\bibfnamefont {S.}~\bibnamefont {Dietrich}}, \bibinfo {author} {\bibfnamefont {C.}~\bibnamefont {Forsythe}}, \bibinfo {author} {\bibfnamefont {T.}~\bibnamefont {Taniguchi}}, \bibinfo {author} {\bibfnamefont {K.}~\bibnamefont {Watanabe}}, \bibinfo {author} {\bibfnamefont {P.}~\bibnamefont {Moon}},\ and\ \bibinfo {author} {\bibfnamefont {C.~R.}\ \bibnamefont {Dean}},\ }\bibfield  {title} {\bibinfo {title} {Anisotropic band flattening in graphene with one-dimensional superlattices},\ }\href@noop {} {\bibfield  {journal} {\bibinfo  {journal} {Nature}\ }\textbf {\bibinfo {volume} {16}},\ \bibinfo {pages} {525} (\bibinfo {year} {2021})}\BibitemShut {NoStop}%
\bibitem [{\citenamefont {Chen}\ \emph {et~al.}(2021)\citenamefont {Chen}, \citenamefont {He}, \citenamefont {Zhang}, \citenamefont {Hsieh}, \citenamefont {Fei}, \citenamefont {Watanabe}, \citenamefont {Taniguchi}, \citenamefont {Cobden}, \citenamefont {Xu}, \citenamefont {Dean},\ and\ \citenamefont {Yankowitz}}]{Chen2021}%
  \BibitemOpen
  \bibfield  {author} {\bibinfo {author} {\bibfnamefont {S.}~\bibnamefont {Chen}}, \bibinfo {author} {\bibfnamefont {M.}~\bibnamefont {He}}, \bibinfo {author} {\bibfnamefont {Y.-H.}\ \bibnamefont {Zhang}}, \bibinfo {author} {\bibfnamefont {V.}~\bibnamefont {Hsieh}}, \bibinfo {author} {\bibfnamefont {Z.}~\bibnamefont {Fei}}, \bibinfo {author} {\bibfnamefont {K.}~\bibnamefont {Watanabe}}, \bibinfo {author} {\bibfnamefont {T.}~\bibnamefont {Taniguchi}}, \bibinfo {author} {\bibfnamefont {D.~H.}\ \bibnamefont {Cobden}}, \bibinfo {author} {\bibfnamefont {X.}~\bibnamefont {Xu}}, \bibinfo {author} {\bibfnamefont {C.~R.}\ \bibnamefont {Dean}},\ and\ \bibinfo {author} {\bibfnamefont {M.}~\bibnamefont {Yankowitz}},\ }\bibfield  {title} {\bibinfo {title} {Electrically tunable correlated and topological states in twisted monolayer–bilayer graphene},\ }\href@noop {} {\bibfield  {journal} {\bibinfo  {journal} {Nature Physics}\ }\textbf {\bibinfo {volume} {17}},\ \bibinfo {pages} {374–380} (\bibinfo {year}
  {2021})}\BibitemShut {NoStop}%
\bibitem [{\citenamefont {Lilly}\ \emph {et~al.}(1999)\citenamefont {Lilly}, \citenamefont {Cooper}, \citenamefont {Eisenstein}, \citenamefont {Pfeiffer},\ and\ \citenamefont {West}}]{Lilly1999}%
  \BibitemOpen
  \bibfield  {author} {\bibinfo {author} {\bibfnamefont {M.~P.}\ \bibnamefont {Lilly}}, \bibinfo {author} {\bibfnamefont {K.~B.}\ \bibnamefont {Cooper}}, \bibinfo {author} {\bibfnamefont {J.~P.}\ \bibnamefont {Eisenstein}}, \bibinfo {author} {\bibfnamefont {L.~N.}\ \bibnamefont {Pfeiffer}},\ and\ \bibinfo {author} {\bibfnamefont {K.~W.}\ \bibnamefont {West}},\ }\bibfield  {title} {\bibinfo {title} {Evidence for an anisotropic state of two-dimensional electrons in high {{Landau}} levels},\ }\href@noop {} {\bibfield  {journal} {\bibinfo  {journal} {Phys. Rev. Lett.}\ }\textbf {\bibinfo {volume} {82}},\ \bibinfo {pages} {394} (\bibinfo {year} {1999})}\BibitemShut {NoStop}%
\bibitem [{\citenamefont {De~Mey}(1983)}]{deMey1983}%
  \BibitemOpen
  \bibfield  {author} {\bibinfo {author} {\bibfnamefont {G.}~\bibnamefont {De~Mey}},\ }\bibfield  {title} {\bibinfo {title} {Potential calculations in {Hall} plates},\ }\href@noop {} {\bibfield  {journal} {\bibinfo  {journal} {Advances in Electronics and Electron Physics}\ }\textbf {\bibinfo {volume} {61}},\ \bibinfo {pages} {1} (\bibinfo {year} {1983})}\BibitemShut {NoStop}%
\bibitem [{\citenamefont {Zhang}\ \emph {et~al.}(2024)\citenamefont {Zhang}, \citenamefont {Lin}, \citenamefont {Chichinadze}, \citenamefont {Wang}, \citenamefont {Watanabe}, \citenamefont {Taniguchi}, \citenamefont {Fu},\ and\ \citenamefont {li}}]{Zhang2024}%
  \BibitemOpen
  \bibfield  {author} {\bibinfo {author} {\bibfnamefont {N.~J.}\ \bibnamefont {Zhang}}, \bibinfo {author} {\bibfnamefont {J.-X.}\ \bibnamefont {Lin}}, \bibinfo {author} {\bibfnamefont {D.~V.}\ \bibnamefont {Chichinadze}}, \bibinfo {author} {\bibfnamefont {Y.}~\bibnamefont {Wang}}, \bibinfo {author} {\bibfnamefont {K.}~\bibnamefont {Watanabe}}, \bibinfo {author} {\bibfnamefont {T.}~\bibnamefont {Taniguchi}}, \bibinfo {author} {\bibfnamefont {L.}~\bibnamefont {Fu}},\ and\ \bibinfo {author} {\bibfnamefont {J.~I.~A.}\ \bibnamefont {li}},\ }\bibfield  {title} {\bibinfo {title} {Angle-resolved transport non-reciprocity and spontaneous symmetry breaking in twisted trilayer graphene},\ }\href@noop {} {\bibfield  {journal} {\bibinfo  {journal} {Nature Materials}\ }\textbf {\bibinfo {volume} {23}},\ \bibinfo {pages} {356} (\bibinfo {year} {2024})}\BibitemShut {NoStop}%
\bibitem [{\citenamefont {Cui}\ \emph {et~al.}(2017)\citenamefont {Cui}, \citenamefont {Shih}, \citenamefont {Jauregui}, \citenamefont {Chae}, \citenamefont {Kim}, \citenamefont {Li}, \citenamefont {Seo}, \citenamefont {Pistunova}, \citenamefont {Yin}, \citenamefont {Park}, \citenamefont {Choi}, \citenamefont {Watanabe}, \citenamefont {Takashi}, \citenamefont {Kim}, \citenamefont {Dean},\ and\ \citenamefont {Hone}}]{Cui2017}%
  \BibitemOpen
  \bibfield  {author} {\bibinfo {author} {\bibfnamefont {X.}~\bibnamefont {Cui}}, \bibinfo {author} {\bibfnamefont {E.-M.}\ \bibnamefont {Shih}}, \bibinfo {author} {\bibfnamefont {L.~A.}\ \bibnamefont {Jauregui}}, \bibinfo {author} {\bibfnamefont {a.~H.}\ \bibnamefont {Chae}}, \bibinfo {author} {\bibfnamefont {Y.~D.}\ \bibnamefont {Kim}}, \bibinfo {author} {\bibfnamefont {B.}~\bibnamefont {Li}}, \bibinfo {author} {\bibfnamefont {D.}~\bibnamefont {Seo}}, \bibinfo {author} {\bibfnamefont {K.}~\bibnamefont {Pistunova}}, \bibinfo {author} {\bibfnamefont {J.}~\bibnamefont {Yin}}, \bibinfo {author} {\bibfnamefont {J.-H.}\ \bibnamefont {Park}}, \bibinfo {author} {\bibfnamefont {Y.~H.}\ \bibnamefont {Choi}, \bibfnamefont {Heon-Jin~Lee}}, \bibinfo {author} {\bibfnamefont {K.~T.}\ \bibnamefont {Watanabe}}, \bibinfo {author} {\bibnamefont {Takashi}}, \bibinfo {author} {\bibfnamefont {P.}~\bibnamefont {Kim}}, \bibinfo {author} {\bibfnamefont {C.~R.}\ \bibnamefont {Dean}},\ and\ \bibinfo {author} {\bibfnamefont {J.~C.}\
  \bibnamefont {Hone}},\ }\bibfield  {title} {\bibinfo {title} {Low-temperature ohmic contact to monolayer {MoS$_2$} by van der {Waals} bonded co/h-bn electrodes},\ }\href@noop {} {\bibfield  {journal} {\bibinfo  {journal} {Nano Letters}\ }\textbf {\bibinfo {volume} {17}},\ \bibinfo {pages} {4781–4786} (\bibinfo {year} {2017})}\BibitemShut {NoStop}%
\bibitem [{\citenamefont {Tomarken}\ \emph {et~al.}(2019)\citenamefont {Tomarken}, \citenamefont {Cao}, \citenamefont {Demir}, \citenamefont {Watanabe}, \citenamefont {Taniguchi}, \citenamefont {Jarillo-Herrero},\ and\ \citenamefont {Ashoori}}]{Tomarken2019}%
  \BibitemOpen
  \bibfield  {author} {\bibinfo {author} {\bibfnamefont {S.}~\bibnamefont {Tomarken}}, \bibinfo {author} {\bibfnamefont {Y.}~\bibnamefont {Cao}}, \bibinfo {author} {\bibfnamefont {A.}~\bibnamefont {Demir}}, \bibinfo {author} {\bibfnamefont {K.}~\bibnamefont {Watanabe}}, \bibinfo {author} {\bibfnamefont {T.}~\bibnamefont {Taniguchi}}, \bibinfo {author} {\bibfnamefont {P.}~\bibnamefont {Jarillo-Herrero}},\ and\ \bibinfo {author} {\bibfnamefont {R.}~\bibnamefont {Ashoori}},\ }\bibfield  {title} {\bibinfo {title} {Electronic compressibility of magic-angle graphene superlattices},\ }\href@noop {} {\bibfield  {journal} {\bibinfo  {journal} {Physical Review Letters}\ }\textbf {\bibinfo {volume} {123}} (\bibinfo {year} {2019})}\BibitemShut {NoStop}%
\bibitem [{\citenamefont {Gustafsson}\ \emph {et~al.}(2019)\citenamefont {Gustafsson}, \citenamefont {Yankowitz}, \citenamefont {Forsythe}, \citenamefont {Rhodes}, \citenamefont {Watanabe}, \citenamefont {Taniguchi}, \citenamefont {Hone}, \citenamefont {Zhu},\ and\ \citenamefont {Dean}}]{Gustafsson2018}%
  \BibitemOpen
  \bibfield  {author} {\bibinfo {author} {\bibfnamefont {M.}~\bibnamefont {Gustafsson}}, \bibinfo {author} {\bibfnamefont {M.}~\bibnamefont {Yankowitz}}, \bibinfo {author} {\bibfnamefont {C.}~\bibnamefont {Forsythe}}, \bibinfo {author} {\bibfnamefont {D.}~\bibnamefont {Rhodes}}, \bibinfo {author} {\bibfnamefont {K.}~\bibnamefont {Watanabe}}, \bibinfo {author} {\bibfnamefont {T.}~\bibnamefont {Taniguchi}}, \bibinfo {author} {\bibfnamefont {J.}~\bibnamefont {Hone}}, \bibinfo {author} {\bibfnamefont {X.}~\bibnamefont {Zhu}},\ and\ \bibinfo {author} {\bibfnamefont {C.}~\bibnamefont {Dean}},\ }\bibfield  {title} {\bibinfo {title} {Ambipolar {{Landau}} levels and strong band-selective carrier interactions in monolayer {WSe$_2$}},\ }\href@noop {} {\bibfield  {journal} {\bibinfo  {journal} {Nature Materials}\ }\textbf {\bibinfo {volume} {17}},\ \bibinfo {pages} {411} (\bibinfo {year} {2019})}\BibitemShut {NoStop}%
\bibitem [{\citenamefont {Jin}\ \emph {et~al.}(2021)\citenamefont {Jin}, \citenamefont {Tau}, \citenamefont {Li}, \citenamefont {Xu}, \citenamefont {Tang}, \citenamefont {Zhu}, \citenamefont {Liu}, \citenamefont {Watanabe}, \citenamefont {Taniguchi}, \citenamefont {Hone}, \citenamefont {Fu}, \citenamefont {Shan},\ and\ \citenamefont {Mak}}]{Jin2021}%
  \BibitemOpen
  \bibfield  {author} {\bibinfo {author} {\bibfnamefont {C.}~\bibnamefont {Jin}}, \bibinfo {author} {\bibfnamefont {Z.}~\bibnamefont {Tau}}, \bibinfo {author} {\bibfnamefont {T.}~\bibnamefont {Li}}, \bibinfo {author} {\bibfnamefont {Y.}~\bibnamefont {Xu}}, \bibinfo {author} {\bibfnamefont {Y.}~\bibnamefont {Tang}}, \bibinfo {author} {\bibfnamefont {J.}~\bibnamefont {Zhu}}, \bibinfo {author} {\bibfnamefont {S.}~\bibnamefont {Liu}}, \bibinfo {author} {\bibfnamefont {K.}~\bibnamefont {Watanabe}}, \bibinfo {author} {\bibfnamefont {T.}~\bibnamefont {Taniguchi}}, \bibinfo {author} {\bibfnamefont {J.~C.}\ \bibnamefont {Hone}}, \bibinfo {author} {\bibfnamefont {L.}~\bibnamefont {Fu}}, \bibinfo {author} {\bibfnamefont {J.}~\bibnamefont {Shan}},\ and\ \bibinfo {author} {\bibfnamefont {K.~F.}\ \bibnamefont {Mak}},\ }\bibfield  {title} {\bibinfo {title} {Stripe phases in {WSe$_2$}/{WS$_2$} moiré superlattices},\ }\href@noop {} {\bibfield  {journal} {\bibinfo  {journal} {Nature Materials}\ }\textbf {\bibinfo {volume}
  {20}},\ \bibinfo {pages} {940} (\bibinfo {year} {2021})}\BibitemShut {NoStop}%
\bibitem [{\citenamefont {Zhu}\ \emph {et~al.}(2010)\citenamefont {Zhu}, \citenamefont {Sambandamurthy}, \citenamefont {Chen}, \citenamefont {Jiang}, \citenamefont {Engel}, \citenamefont {Tsui}, \citenamefont {Pfeiffer},\ and\ \citenamefont {West}}]{Zhu2010}%
  \BibitemOpen
  \bibfield  {author} {\bibinfo {author} {\bibfnamefont {H.}~\bibnamefont {Zhu}}, \bibinfo {author} {\bibfnamefont {G.}~\bibnamefont {Sambandamurthy}}, \bibinfo {author} {\bibfnamefont {Y.~P.}\ \bibnamefont {Chen}}, \bibinfo {author} {\bibfnamefont {P.}~\bibnamefont {Jiang}}, \bibinfo {author} {\bibfnamefont {L.~W.}\ \bibnamefont {Engel}}, \bibinfo {author} {\bibfnamefont {D.~C.}\ \bibnamefont {Tsui}}, \bibinfo {author} {\bibfnamefont {L.~N.}\ \bibnamefont {Pfeiffer}},\ and\ \bibinfo {author} {\bibfnamefont {K.~W.}\ \bibnamefont {West}},\ }\bibfield  {title} {\bibinfo {title} {Pinning-mode resonance of a skyrme crystal near {Landau}-level filling factor $\nu$=1},\ }\href@noop {} {\bibfield  {journal} {\bibinfo  {journal} {Physical Review Letters}\ }\textbf {\bibinfo {volume} {104}},\ \bibinfo {pages} {226801} (\bibinfo {year} {2010})}\BibitemShut {NoStop}%
\bibitem [{\citenamefont {Hatke}\ \emph {et~al.}(2014)\citenamefont {Hatke}, \citenamefont {Liu}, \citenamefont {Magill}, \citenamefont {Moon}, \citenamefont {Engel}, \citenamefont {Shayegan}, \citenamefont {Pfeiffer}, \citenamefont {West},\ and\ \citenamefont {Baldwin}}]{Hatke2014}%
  \BibitemOpen
  \bibfield  {author} {\bibinfo {author} {\bibfnamefont {A.~T.}\ \bibnamefont {Hatke}}, \bibinfo {author} {\bibfnamefont {Y.}~\bibnamefont {Liu}}, \bibinfo {author} {\bibfnamefont {B.~A.}\ \bibnamefont {Magill}}, \bibinfo {author} {\bibfnamefont {B.~H.}\ \bibnamefont {Moon}}, \bibinfo {author} {\bibfnamefont {L.~W.}\ \bibnamefont {Engel}}, \bibinfo {author} {\bibfnamefont {M.}~\bibnamefont {Shayegan}}, \bibinfo {author} {\bibfnamefont {L.~N.}\ \bibnamefont {Pfeiffer}}, \bibinfo {author} {\bibfnamefont {K.~W.}\ \bibnamefont {West}},\ and\ \bibinfo {author} {\bibfnamefont {K.~W.}\ \bibnamefont {Baldwin}},\ }\bibfield  {title} {\bibinfo {title} {Microwave spectroscopic observation of distinct electron solid phases in wide quantum wells},\ }\href@noop {} {\bibfield  {journal} {\bibinfo  {journal} {Nature Communications}\ }\textbf {\bibinfo {volume} {5}},\ \bibinfo {pages} {4154} (\bibinfo {year} {2014})}\BibitemShut {NoStop}%
\bibitem [{\citenamefont {Chen}\ \emph {et~al.}(2006)\citenamefont {Chen}, \citenamefont {Sambandamurthy}, \citenamefont {Wang}, \citenamefont {Lewis}, \citenamefont {Engel}, \citenamefont {Tsui}, \citenamefont {Ye}, \citenamefont {Pfeiffer},\ and\ \citenamefont {West}}]{Chen2006}%
  \BibitemOpen
  \bibfield  {author} {\bibinfo {author} {\bibfnamefont {Y.~P.}\ \bibnamefont {Chen}}, \bibinfo {author} {\bibfnamefont {G.}~\bibnamefont {Sambandamurthy}}, \bibinfo {author} {\bibfnamefont {Z.~H.}\ \bibnamefont {Wang}}, \bibinfo {author} {\bibfnamefont {R.~M.}\ \bibnamefont {Lewis}}, \bibinfo {author} {\bibfnamefont {L.~W.}\ \bibnamefont {Engel}}, \bibinfo {author} {\bibfnamefont {D.~C.}\ \bibnamefont {Tsui}}, \bibinfo {author} {\bibfnamefont {P.~D.}\ \bibnamefont {Ye}}, \bibinfo {author} {\bibfnamefont {L.~N.}\ \bibnamefont {Pfeiffer}},\ and\ \bibinfo {author} {\bibfnamefont {K.~W.}\ \bibnamefont {West}},\ }\bibfield  {title} {\bibinfo {title} {Melting of a 2d quantum electron solid in high magnetic field},\ }\href@noop {} {\bibfield  {journal} {\bibinfo  {journal} {Nature Physics}\ }\textbf {\bibinfo {volume} {2}},\ \bibinfo {pages} {452} (\bibinfo {year} {2006})}\BibitemShut {NoStop}%
\bibitem [{\citenamefont {Andrei}\ \emph {et~al.}(1988)\citenamefont {Andrei}, \citenamefont {Deville}, \citenamefont {Glattli}, \citenamefont {Williams}, \citenamefont {Paris},\ and\ \citenamefont {Etienne}}]{Andrei1988}%
  \BibitemOpen
  \bibfield  {author} {\bibinfo {author} {\bibfnamefont {E.~Y.}\ \bibnamefont {Andrei}}, \bibinfo {author} {\bibfnamefont {G.}~\bibnamefont {Deville}}, \bibinfo {author} {\bibfnamefont {D.~C.}\ \bibnamefont {Glattli}}, \bibinfo {author} {\bibfnamefont {F.~I.~B.}\ \bibnamefont {Williams}}, \bibinfo {author} {\bibfnamefont {E.}~\bibnamefont {Paris}},\ and\ \bibinfo {author} {\bibfnamefont {B.}~\bibnamefont {Etienne}},\ }\bibfield  {title} {\bibinfo {title} {Observation of a magnetically induced {Wigner} solid},\ }\href@noop {} {\bibfield  {journal} {\bibinfo  {journal} {Phys. Rev. Lett.}\ }\textbf {\bibinfo {volume} {60}},\ \bibinfo {pages} {1926} (\bibinfo {year} {1988})}\BibitemShut {NoStop}%
\bibitem [{\citenamefont {Liu}\ \emph {et~al.}(2012)\citenamefont {Liu}, \citenamefont {Pappas}, \citenamefont {Shayegan}, \citenamefont {Pfeiffer}, \citenamefont {West},\ and\ \citenamefont {Baldwin}}]{Liu2012}%
  \BibitemOpen
  \bibfield  {author} {\bibinfo {author} {\bibfnamefont {Y.}~\bibnamefont {Liu}}, \bibinfo {author} {\bibfnamefont {C.~G.}\ \bibnamefont {Pappas}}, \bibinfo {author} {\bibfnamefont {M.}~\bibnamefont {Shayegan}}, \bibinfo {author} {\bibfnamefont {L.~N.}\ \bibnamefont {Pfeiffer}}, \bibinfo {author} {\bibfnamefont {K.~W.}\ \bibnamefont {West}},\ and\ \bibinfo {author} {\bibfnamefont {K.~W.}\ \bibnamefont {Baldwin}},\ }\bibfield  {title} {\bibinfo {title} {Observation of reentrant integer quantum {Hall} states in the lowest {Landau} level},\ }\href@noop {} {\bibfield  {journal} {\bibinfo  {journal} {Physical Review Letters}\ }\textbf {\bibinfo {volume} {109}},\ \bibinfo {pages} {036801} (\bibinfo {year} {2012})}\BibitemShut {NoStop}%
\bibitem [{\citenamefont {Chen}\ \emph {et~al.}(2019)\citenamefont {Chen}, \citenamefont {Ribeiro-Palau}, \citenamefont {Yang}, \citenamefont {Watanabe}, \citenamefont {Taniguchi}, \citenamefont {Hone}, \citenamefont {Goerbig},\ and\ \citenamefont {Dean}}]{Chen2019}%
  \BibitemOpen
  \bibfield  {author} {\bibinfo {author} {\bibfnamefont {S.}~\bibnamefont {Chen}}, \bibinfo {author} {\bibfnamefont {R.}~\bibnamefont {Ribeiro-Palau}}, \bibinfo {author} {\bibfnamefont {K.}~\bibnamefont {Yang}}, \bibinfo {author} {\bibfnamefont {K.}~\bibnamefont {Watanabe}}, \bibinfo {author} {\bibfnamefont {T.}~\bibnamefont {Taniguchi}}, \bibinfo {author} {\bibfnamefont {J.}~\bibnamefont {Hone}}, \bibinfo {author} {\bibfnamefont {M.~O.}\ \bibnamefont {Goerbig}},\ and\ \bibinfo {author} {\bibfnamefont {C.~R.}\ \bibnamefont {Dean}},\ }\bibfield  {title} {\bibinfo {title} {Competing fractional quantum {Hall} and electron solid phases in graphene},\ }\href@noop {} {\bibfield  {journal} {\bibinfo  {journal} {Physical Review Letters}\ }\textbf {\bibinfo {volume} {122}},\ \bibinfo {pages} {026802} (\bibinfo {year} {2019})}\BibitemShut {NoStop}%
\end{thebibliography}

\providecommand{\noopsort}[1]{}\providecommand{\singleletter}[1]{#1}%

\end{document}